# FAZENDO 3D COM UMA CÂMERA SÓ
## (SINGLE CAMERA 3D)

José J. Lunazzi
Universidade Estadual de Campinas, Instituto de Física
lunazzi  xatx ifi.unicamp.br

**Abstract**
A simple system to make stereo photography or videos based in just two mirrors was made in 1988 and recently adapted to a digital camera setup. Possibilities in the use of computer software to make anaglyphic stereo are also reported.

**Resumo**
Um sistema simples para fazer fotografia ou vídeos em estéreo baseado em dois espelhos que dividem o campo da imagem foi criado no ano 1989, e recentemente adaptado para câmera digital. Reporta-se algumas possibilidades para se fazer estéreo 3D bicolor com programas de computador.

**Palavras-chave:** Imagem 3D, Imagem tridimensional, fotografia 3D, vídeo 3D, estereoscopia, anaglifos

## 1. Introdução

Algumas pessoas, sobretudo os mais jóvens, podem achar que o 3D estéreo é uma novidade, trazida pelo cinema e ainda mais pelo filme Avatar. Mas a estereofotografia, ou fotografia estéreo, foi inventada somente alguns anos depois que a fotografia foi conhecida, começando pelo estereoscópio de Charles Wheatstone em 1838(1), o bem conhecido pesquisador que nos deixou importantes trabalhos sobre eletricidade (como o famoso "ponte de Wheatstone"), e aperfeiçoado por David Brewster(2), alguém bem conhecido também dos estudantes de física, pelas suas contribuições na óptica (como o famoso "ângulo de Brewster").

Eu tinha 14 anos de idade e fazia fotos estéreo de objetos e até de seres vivos imóveis. Colocava minha câmera na posição do olho esquerdo e expunha, passando logo sem deslocar o corpo para a posição do olho direito, carregando o filme para a tomada seguinte, e expondo novamente. Era necessário que os objetos ou seres ficassem perfeitamente imóveis frente à câmera. Fazendo cópia de contato, que consiste em colocar um filme virgem baixo o já revelado para expor a sombra e passar de negativo a positivo, técnica inventada em Campinas por Hércules Florence(3), dispunha de um par para montar na moldura de papelão que encaixava no visualizador de duas lentes, algo relativamente comum na época pois acostumava-se registrar casamentos, por exemplo, em 3D com esse visor.

No ano 1988 tinha a possibilidade de projetar vídeo 3D sem óculos pela tela holográfica que acabava de inventar(4). Não tinha como usar duas câmeras e sincronizar as tomadas, um problema que ainda hoje existe. Foi quando pensei em dividir o campo por meio de espelhos, pois existiam referências a separadores de imagem por meio de prismas. Apenas que a divisão simétrica necessitava de um conjunto de prismas equivalente a quatro espelhos, ou quatro espelhos mesmo. Isso complicava a construção. Passei a dividir por meio de dois espelhos, um próximo da lente e outro à distância correspondente a uma visão binocular. Colocando a filmadora VHS, com a lente zoom como teleobjetiva para, maximizando a distância, reduzir a diferença de tamanho e tolerância focal de cada meio campo. A imagem que chegava do espelho mais afastado era algo como um 10% menor que a que chegava pelo primeiro espelho. Filmei assim cenas de mim mesmo em 1988 e no ano seguinte cinquenta minutos de mímica do talentoso ator Luiz Otávio Burnier, da UNICAMP. Junto a outros trabalhos de holoprojeção e fotografias artísticas baseadas em hologramas apresentei o trabalho na Semana de Arte da UNICAMP de 1989, grande evento realizado no Instituto de Artes da UNICAMP. Não existiam, ao menos no Brasil, projetores de vídeo na época e o que fiz foi usar uma TV pequena com o brilho ao máximo e duas lentes na frente para convergir as duas partes do par estéreo sobre tela holográfica de 15 cm x 30 cm(5)(6). O público conheceu a TV estéreo sem óculos pela mímica de um ator e considerei hoje que aquilo tinha algum valor histórico e também prático, mesmo que estejam hoje surgindo câmeras profissionais e serviços no Brasil(7)(8) , por isso resolvi aproveitar as facilidades dos novos programas de edição digital de vídeo para endireitar e redimensionar o arquivo, projetando sobre o novo sistema de TV estéreo com tela holográfica que acabamos de montar com meus alunos que é maior, mais colorido e brilhante(9)(10)(11) que o de 1989. Estamos mostrando ele ao público duas vezes por semana e já foi aproveitado por alunos do curso de mímica, porque o ator que representou é falecido há quinze anos. Passo a descrever o sistema de tomada com somente dois espelhos porque ainda pode ser útil em alguns casos pois, além de dispensar a segunda câmera, garante a sincronização em cenas de ação.

**2. Sistema simples para fazer fotos e vídeos 3D baseado em dois espelhos.**

Retomamos a ideia do antigo sistema, que incluía deslocadores lineares e angulares para os espelhos, e fizemos um simplificado e menor para uma simples câmera de vídeo atual. Não há necessidade de se colocar a câmera invertida como era quando devíamos compensar a inversão pelas lentes projetoras. Se fosse preciso, inversões podem ser realizadas na fase da edição digital. Usamos uma caixa de madeira quase cheia de areia e dois espelhos. A limitação deste sistema está em que troca o formato paisagem pelo retrato, usando metade da resolução. Para conservar o formato paisagem seria preciso um sistema mais sofisticado, com mais dois ou três espelhos. As dimensões da caixa são 31 x 14 cm e a câmera fica apoiada em um extremo entanto que os espelhos, de 120 x 30 x 2 mm$^3$ o primeiro e

150 x 70 x 2 mm³ o segundo são colocados na distância de 85 mm entre centros para um objeto a registrar na distância de 2,5 m. Colocando a referência nas bordas dos espelhos, indicamos que a distância da câmera ao primeiro espelho é de 5,5 cm e a distancia da câmera ao segundo espelho é de 13 cm. A Figura 1 mostra o sistema montado.

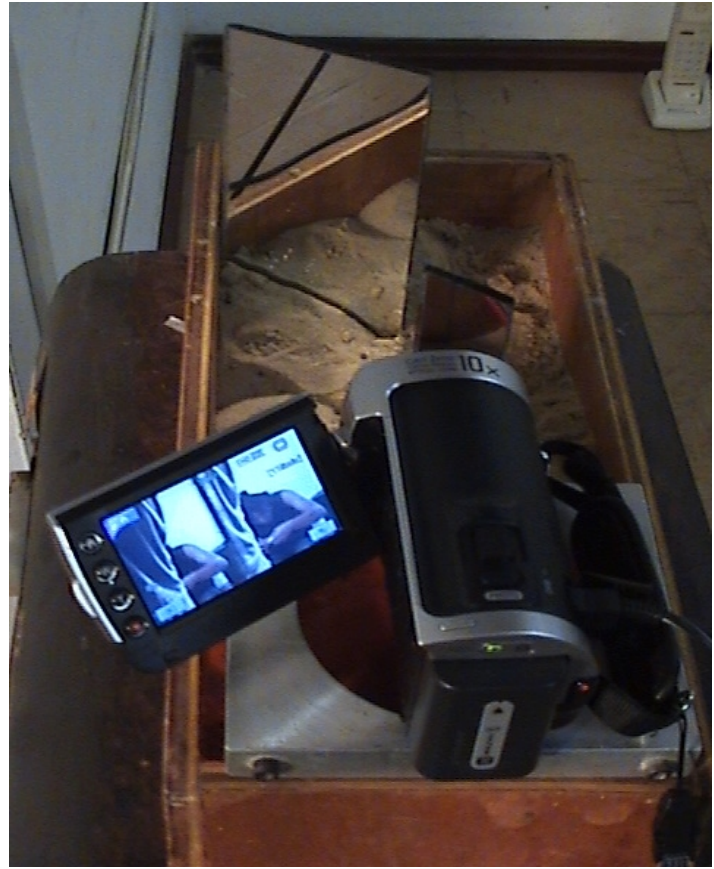

*Figura 1: Sistema de câmera e dois espelhos montado em uma caixa de areia. Note a divisão vertical de campos na tela.*

Os espelhos devem ser alinhados cuidando de que o campo fique dividido em duas partes iguais e centralizando o sujeito a registrar observando um ponto de referência. Se for uma pessoa, por exemplo, o nariz dela deve aparecer no centro e na mesma altura nos dois campos de imagem, como na Figura 2.

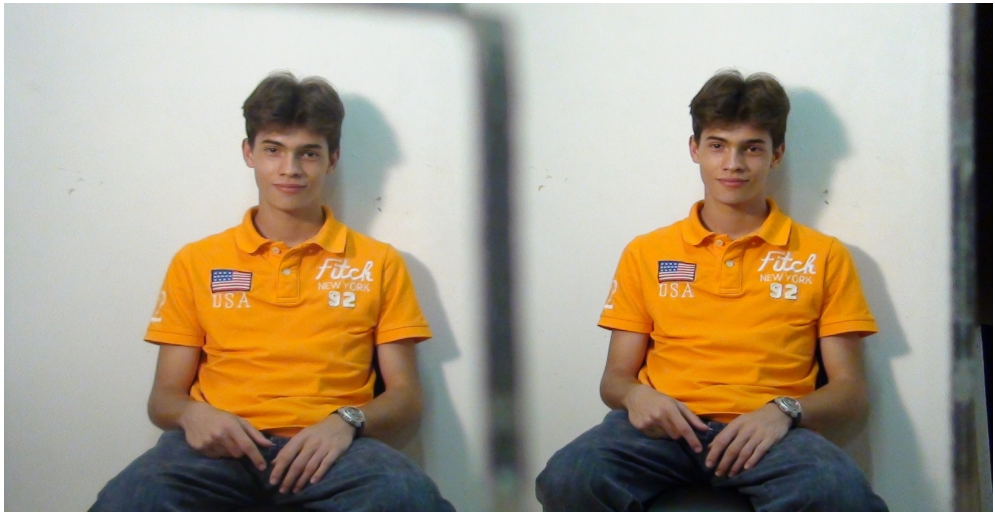

*Figura 2: Par estéreo paralelo, com a vista esquerda a esquerda, e direita a direita. Note que a altura dos elementos está alinhada horizontalmente. A cena foi espelhada, ou seja, invertida digitalmente esquerda-direita para dar a relação correta, como se vê na leitura de letras.*

O sistema é assimétrico porque os raios da luz captada pelo primeiro espelho percorrem uma distância menor que os que são imageados através do segundo espelho. Isto resulta em uma imagem algo maior para o primeiro, da ordem de 2%, que pode ser corrigida digitalmente, e em uma diferença na focalização, seja esta fixa ou automática, que tem pouca importância pois foi observado que o cérebro pode recompor a nitidez com uma das cenas enquanto usa a outra para determinar a profundidade(12). Deve-se ter cuidado de não tentar forçar uma aproximação óptica por variação da distância focal da lente (o chamado "zoom") porque podemos perder a base de triangulação, o elemento que aproxima o resultado de nossa visão binocular. Pense que a separação com que vemos uma cena é a de nossos olhos, algo como 65 mm, e nossos olhos não tem zoom.

## 3. Processamento digital das fotos obtidas

Usamos o programa livre GIMP que pode ser obtido pela internet (www.gimp.org), existente em versão para Linux e para Windows, mas é claro que recomendamos trabalhar em Linux por economia, para evitar a pirataria, e pelo conhecimento total que pode-se ter da operação dos programas em nosso micro. Além de não criar preocupação alguma com vírus e, consequentemente, com programas anti-vírus. Entendamos primeiramente o princípio do que procuramos para melhor lembrar depois das operações: Para barrar a cada olho uma das duas imagens os filtros dos óculos deixam passar uma cor nos extremos do espectro visível, ou seja o vermelho (sempre) por um filtro e o verde ou o azul pelo outro. A escolha de verde ou azul para o filtro direito altera pouco o resultado e não merece uma discussão mais detalhada neste artigo, podemos escolher a cor que quisermos. Trabalhamos com o sistema de tricromia que controla os subpixels da tela, vermelho verde e azul, tecnicamente conhecidos como RGB. R = vermelho, G = verde, B = azul. O programa deve então estar trabalhando as cores nessa modalidade de três canais de base, a RGB. A maneira mais simples então é fazer com que uma das fotografias fique apenas no canal vermelho e a outra no azul. O verde, sendo um comprimento de onda intermediário, está mais próximo do vermelho e por isso nunca é escolhido para substituir ao vermelho como filtro para o olho esquerdo. Mas no entanto, o canal verde pode ser adicionado para ser visto pelo olho direito e, embora faltem essas cores no olho esquerdo, o cérebro será capaz de compor uma visualização de cor bastante próxima da completa deixando a cena bem mais agradável, próxima do que seria poder fazer funcionar com óculos polarizados.

O exemplo que damos foi realizado em Ubuntu usando o GIMP 2.7.2 .

1) Devemos primeiramente separar o par em uma imagem esquerda e outra direita. Para isso abrimos o mesmo (Aplicativos, Gráficos, GIMP) e colocamos a opção Ferramentas, Ferramentas de Transformação, Espelhar, correspondente ao atalho SHIFT f, para termos a orientação correta das imagens onde palavras, por exemplo, possam ser lidas.
2) Depois selecionamos Ferramentas, Ferramentas de Transformação, Cortar, que pode também ser ativada pelo atalho SHIFT+c , e selecionamos cuidadosamente o quadro esquerdo, como mostra a Figura 3.

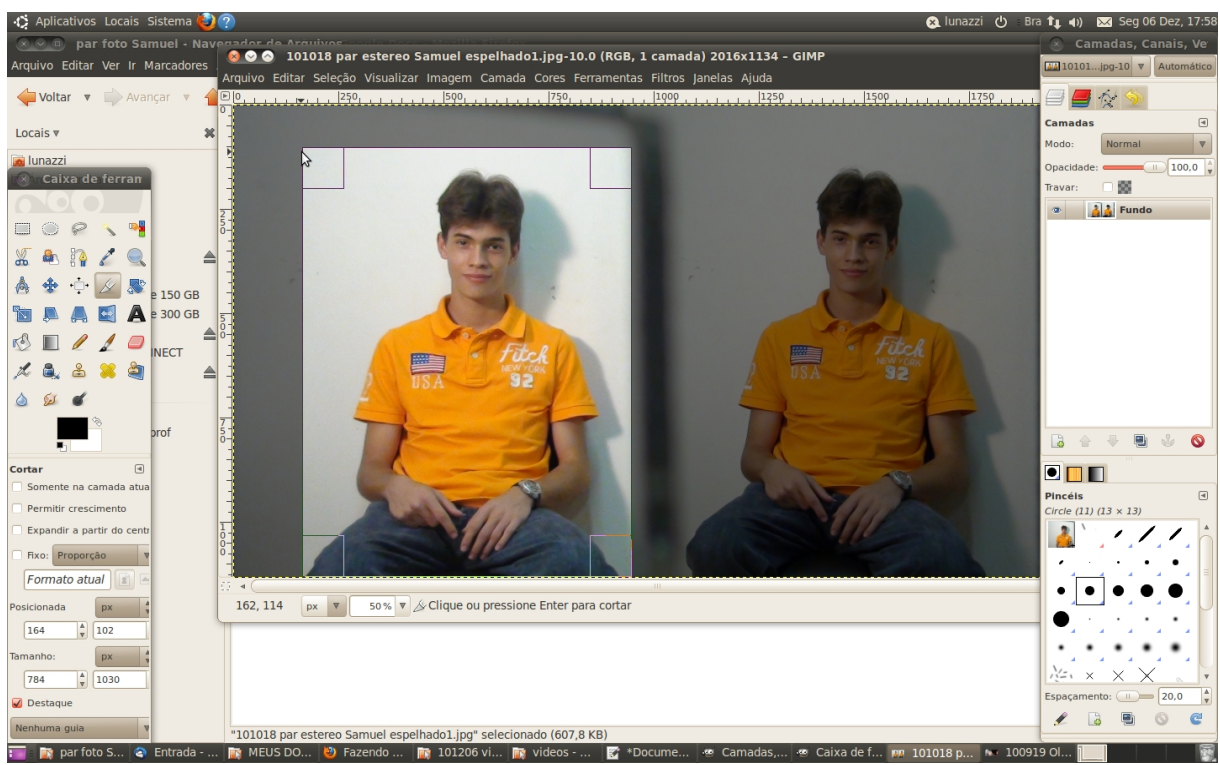

*Figura 3: Seleção do quadro esquerdo a partir do par estéreo já espelhado.*

Devemos tomar a base que nos permita eliminar elementos indesejáveis, como a borda dos espelhos, por exemplo, sendo esta a mesma que vamos utilizar para depois selecionar a cena direita. A altura também. Note-se que a cena deve evitar ao máximo ter elementos importantes na zona de corte, sobretudo nas laterais. Perceba que o que entrar justamente no corte de uma cena do par, pode não estar presente na outra cena dele. É impossível por causa da mesma tridimensionalidade que as duas cenas contenham todos os mesmos elementos, e o cérebro logo percebe e destaca essa diferença. Uma vez feita a seleção, o recorte acontece facilmente clicando dentro da área selecionada. Cuide de salvar o resultado com o nome "cena esquerda" ou equivalente, abra novamente o par e selecione, salve e feche a cena direita.

3) Tendo observado diferença no aumento das imagens, o que pode ser feito medindo em cada uma das cenas dois pontos semelhantes afastados verticalmente o mais possível, calcule o percentual e corrija essa diferença na cena direita usando Ferramentas, Ferramentas de Transformação, Redimensionar (SHIFT t). Selecione para alterar por porcentagem, passe o mouse no valor de largura e coloque o percentual que deseja alterar, por exemplo, 102%, logo clique no ícone em forma de elos de corrente a direita e confira que foi alterado proporcionalmente. Clique em Redimensionar, aplique o mesmo recorte padrão que já usou antes nessa figura que está agora maior, e salve.

4) Abra as duas partes agora usando a opção Abrir, Abrir como camadas (CTRL+ALT+o), selecione os dois arquivos com o mouse segurando a tecla CTRL e clique em "OK". As duas partes estarão sobrepostas mas vai ter na caixa de ferramentas de camadas uma miniatura de cada uma, sendo que uma delas pode estar sendo chamada de "Fundo". Se a janela para controle de camadas não aparecer, colocá-la indo em "Janelas", "Diálogos de encaixe", "Camadas" ou, simplesmente, operando com CTRL l.

5) Desative a cena esquerda clicando no pequeno olho que aparece na esquerda da miniatura e vai em "Cor", "Níveis". Na janela de níveis selecione o canal vermelho e leve o nível da barra de saída de 255 para 0. Assim você vê a cena esverdeada, pois tem azul e verde, mas não o vermelho. Olhando pelo filtro vermelho, aparece preta, pelo azul, normal.

6) Agora troque a seleção de cena, desativando o pequeno ícone de olho e ativando-o na outra, e faça o mesmo com a cena esquerda, mas desta vez anulando os canais de verde e azul, com o que a cena vai ficar vermelha, e preta quando vista pelo filtro azul (não deve haver imagens parasitas nesses escurecimentos) .

7) Agora faça a sobreposição de duas imagens em uma, no painel de camadas em "modo", pase de "Normal" para "Esconder" ("screen") ou "Adição" ("add").. Deve acontecer a fusão das duas cenas em uma só, com silueta vermelha de um lado e azul do outro, como na Figura 4.

8) Já pode ver o resultado usando os óculos bicolor, para salvá-lo deve-o exportar ao formato que quiser, preferentemente o png por ser o mais livre, ou então jpg, etc.

A base escolhida para realizar o corte estaria, idealmente, centrada no elemento principal da cena. Vamos supor que esse fosse o nariz do sujeito. Centrando ela nos dois cortes, o nariz vai aparecer na visão 3D na posição da tela, quer dizer, sem profundidade, e a cabeça por trás enquanto as pernas serão vistas pela frente. É possível deslocar essa posição para o conjunto fazendo o corte estar mais a direita ou esquerda do nariz em uma das vistas. Experimente!

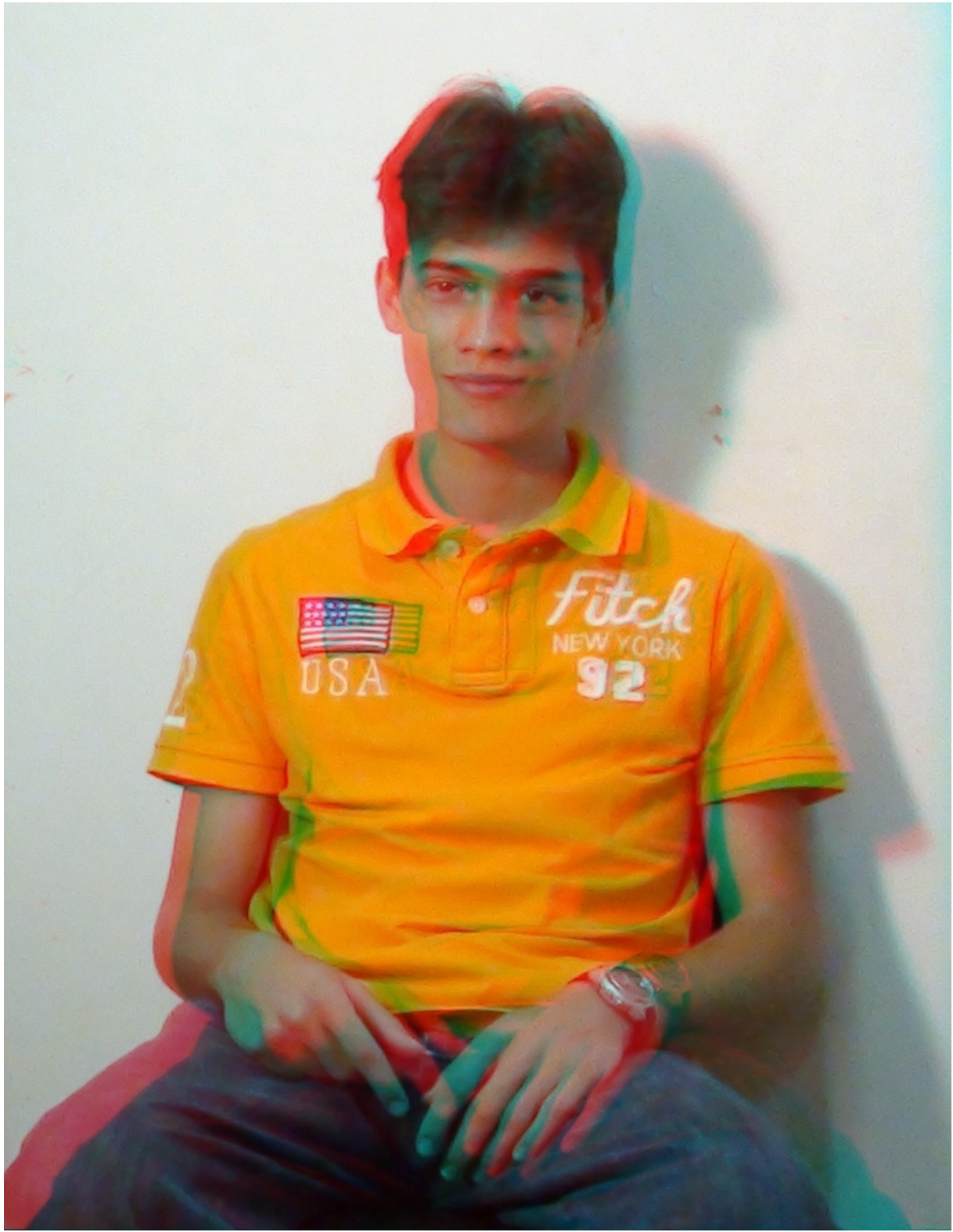

**Fig.4: Imagem estereoscópica final.**

## 4. Processamento digital dos vídeos obtidos

O processamento digital nos vídeos segue exatamente os mesmos lineamentos que no caso da fotografia, mas não podemos contar com programas livres para isso. Os que conheço são proprietários e caros. Temos usado o SONY VEGAS, mas seguramente o Adobe Premiere também consegue realizar a mesma tarefa: recortar os lados do campo, separar os canais de cores de tricromia, e sobrepor os dois campos. Não farei a descrição detalhada neste artigo porque a edição de vídeo é tarefa mais sofisticada do que o nível que pretendo para ele. Apenas cito um breve exemplo que realizamos e que podemos encontrar na internet(13).

## 5. Aperfeiçõamentos possíveis:

A presença de vidro no caminho óptico gera astigmatismo e pode ser eliminada usando vidros com espelhamento na primeira superfície. O processamento digital talvez poderia ser simplificado construindo um programa específico. Estamos trabalhando nisso.

## 6. Realização e impressão de 3D pelo computador

Desde que as telas de computador são coloridas, e faz disto uns vinte anos, que é possível ver e criar cenas em 3D(14). O termo "imagens em 3D" que correspondia à visão por meio de óculos foi copiado das fotos e cinema que existiam em estereoscopia e aplicado a programas de computador que oferecem imagens 2D em perspectiva. É possível com esses programas criar imagens de 3D verdadeiro, basta com ter duas perspectivas correspondentes aos pontos de vista virtual de cada um dos olhos. Separando e recombinando em vermelho e verde, ou vermelho e azul, como na Figura 5. É possível até imprimir e, com sorte, como já tivemos, inclusive em camisetas, o desenho impresso dá uma reconstrução bastante razoável da imagem 3D, como o trabalho de 1995 realizado por Fábio Barros e que mostramos nas Figuras 5 e 6. A Fig. 5 vista por leitura do artigo na tela do computador não passa pelo efeito de tintas, a Fig. 6 inclui esse efeito.

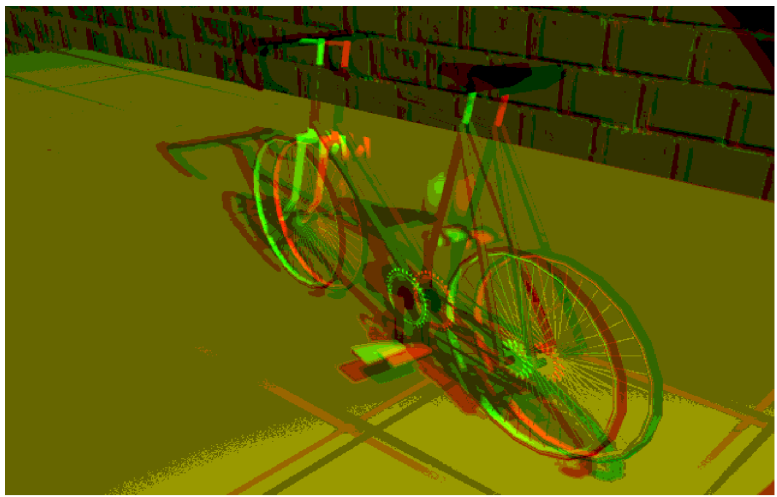

**Fig. 5: acima: Desenho pelo computador de uma bicicleta. Note que o guidão e o assento, em preto, geram erro e ficam em 2D.**

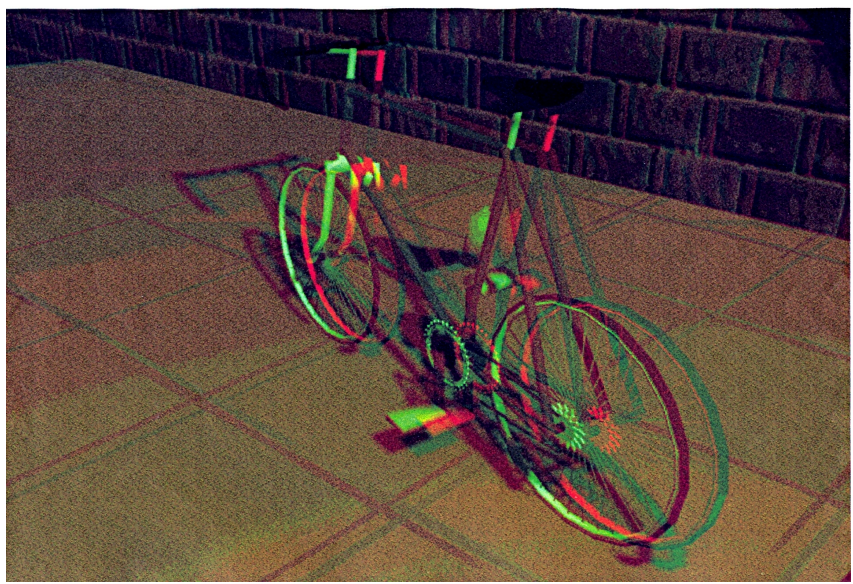

*Fig. 6 Digitalização da imagem da Fig. 4 após ser impressa em papel.*

Animações também podem ser realizadas(15) usando o formato gif. O mesmo programa GIMP que citamos serve para isso, tem um tutorial na internet auxiliando(16). O aluno Cláudio Massao(17) fez uma bonita animação para usar quando chegava a época da copa do Mundo de 1994, e mostrava uma bola entrando pelo gol (Figura 7).

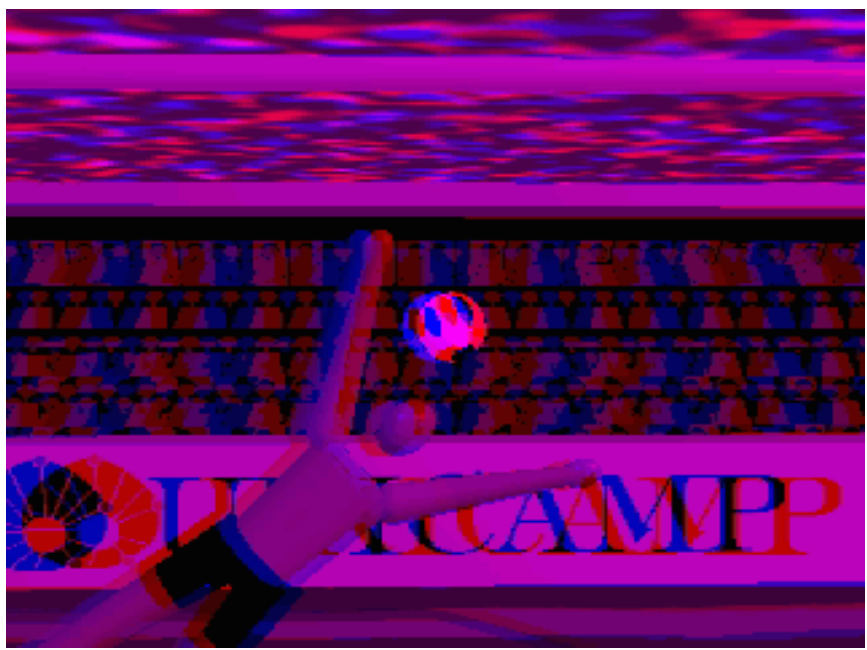

*Figura 7. Quadro da animação anaglífica "Goleiro"*

Esse exemplo serve muito bem para caracterizar a importância da TV 3D no esporte, porque sem a terceira dimensão resulta muito impreciso determinar se a bola atingiu ou não a posição do goleiro.

**7. Construção dos óculos**

O nome da técnica de estereoscopia por separação e filtragem de cores é "anaglífica", mas é um termo que vem do grego e que pouco contribui para a compreensão, prefiro chamá-la de "bicolor", embora as duas cores correspondam apenas aos óculos no caso de usarmos um terceiro canal para cor de imagem. É um exemplo da desinformação que a mídia pode criar o fato de que há tanto tempo que é possível ver e realizar 3D bicolor e no entanto nunca foi comentado isto ao público. Grandes jornais de São Paulo como a Folha e o Estado tem lançado há muitos anos uma ou até duas edições acompanhando os óculos, mas não é dada a indicação de onde encontrar imagem pela internet para aproveitá-los. Revistas de divulgação científica também não o tem feito. Não se consegue comprar óculos facilmente no Brasil, a venda no varejo aparece apenas na internet, e não é muito frequente. Aparecem na internet muitas páginas indicando ser fácil construí-los, mas é errado dizer que podemos usar papel celofane. A qualidade desse material no Brasil é muito baixa, notadamente nas cores verde e azul. O que pode ser utilizado sim é o acetato ou gelatina, que temos conseguido em lojas que trabalham com iluminação para shows ou comprando pela internet. O modelo para recortar os óculos que indico na figura 8 tem campo maior, espaço maior nos retángulos do que os que são indicados na internet, que mais parecem com os óculos comerciais. A recomendação importante (fora, claro, a norma de colocar o vermelho para o olho esquerdo, e o azul para o direito) é a de cortar e colar com extrema limpeza, o menor resto de cola resultaria intolerável. A questão da higiene no uso de óculos por várias pessoas é algo a considerar, nos cinemas do Brasil está-se colocando normas para que a limpeza dos mesmos seja obrigatória. Os problemas relatados, no entanto, não são o de doenças graves, apenas temporárias.

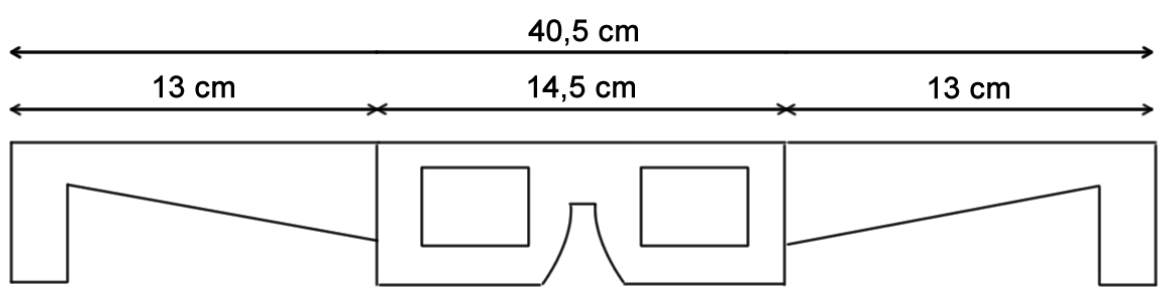

*Fig. 8: formato de óculos recomendado para recorte.*

**8. Observações**

Não porque a montagem de um sistema 3D possa parecer simples deve-se achar que fazer 3D é fácil: as condições para a tomada envolvem uma escolha cuidadosa das distâncias de maneira que a perspectiva seja próxima da de uma situação binocular. Esta escolha criteriosa pode ser feita por meio de cálculos mas necessita também de uma sensibilidade extrema na observação da imagem, incluindo a eliminação ou redução de presença de elementos que ficam à direita e esquerda da imagem porque estes somente aparecem em uma das duas vistas. Qualquer erro em 3D resulta amplificado na comparação que o cérebro faz entre o que chega a cada instante nos olhos. A sensibilidade a diferenças é tanta que os astrónomos sempre utilizaram a montagem em um estereoscópio de duas vistas de uma mesma região do céu registradas em sequencia para detetar a presença de cometas. Mais simplesmente, é possível resolver o conhecido "Jogo dos sete erros" simplesmente cruzando os olhos e sobrepondo as duas cenas (me desculpem os autores destes jogos pela revelação). É pelas falhas que muitos, a maioria talvez, dos trabalhos que aparecem em foto ou vídeo desestimulam ao público a querer ver em estéreo. Existem, por outro lado, muitas possibilidades para se ver 3D estéreo na internet(18), e isto acontece há vinte anos. Mais exatamente desde que os monitores de computador incorporaram a tecnologia da cor, que vinha separada em subpixels com a base vermelho, verde e azul, respeitada pelos programas de criação e reprodução de imagens e figuras, algo que a TV a cores não fazia. Alguns sistemas de visão atuais até dispensam o uso de óculos especiais exigindo que os olhos tomem posições bem determinadas, mas é mais confortável usar os clássicos óculos bicolor, seja como vermelho e verde ou vermelho e azul (outras possibilidades existem). Considero isto um excelente exemplo de como o público fica desinformado pela falta de acesso que temos aos grandes meios de divulgação: dois dos mais importantes jornais do pais e várias revistas tem feito edições incluindo os óculos bicolor, mas nunca foi dito que estes podiam ser utilizados para uso no computador e na internet. Na UNICAMP realizamos fotografias e desenhos desde 1991 e tentamos por todos os meios fazer saber desta possibilidade, orientando inclusive para a construção correta dos óculos.

outorgadas aos alunos Brenno Pereira Machado e Vinícius Emanuel Ares que colaboraram na edição de fotografias e vídeo e na construção de óculos bi-color, respectivamente. Também, ao serviço repositório de manuscritos, pre-prints e artigos da Cornell University www.arxiv.org que permite em 48 h ter o registro de autoria de um trabalho(19).

**Referências:**

1) Wheatstone criou o estereoscópio e, consequentemente, a estereoscopia.
http://en.wikipedia.org/wiki/Stereopsis
2) Brewster usava fotografias e duas lentes para que cada vista fosse enxergada por cada olho separadamente. Isto está muito bem documentado na internet:
http://en.wikipedia.org/wiki/David_Brewster
3) Kossoy, B., *"Hercules Florence, A Descoberta Isolada da Fotografia no Brasil"*, São Paulo, Duas Cidades, 1980. O livro afirma incorretamente que a fotografia foi descoberta em Campinas-SP-BR em 1833. Entendo que não foi um resultado atingido pois não gerou um registro duradouro por técnica original, mas sim o processo de conversão negativo-positivo.
4) J. J. Lunazzi, patente francesa *"Procédé et dispositif pour projeter et observer des images differenciees ou stereoscopiques, dessins, photographies, films cinematographiques ou video"*, INPIFR, Nº 8907241, 1992.
5) J.J. Lunazzi, *"UNICAMP dá início à corrida pelo cinema holográfico"*, Jornal "Folha de São Paulo", 12 de maio de 1989, Caderno "Ciência", página G6.
6) *"Nova Dimensão na Tela de TV",* Revista Superinteressante Ano 4, N2, fevereiro de 1990.
7) Firma brasileira Photon3D, http://www.photon3d.com.br/ Recomendo ver a descrição de diversos sistemas para 3D estéreo alí colocados.
8) Mirela Tavares, sítio "3DHub" mantido pela jornalista, www.3dhub.com.br Podem se encontrar nele exemplos de fotos e algumas matérias do autor.
9) J.J. Lunazzi, D.S.F. Magalhães, R.F. Serra, Opt. Eng. **48**, 9, pp. 085902-085902-5 (2009), *"Construction of white-light holographic screens"*, Opt. Eng., Vol. 48, 095802 (2009); doi:10.1117/1.3231503 http://arxiv.org/pdf/1004.4232
10) J.J. Lunazzi, R.L. Serra, D.S.F. Magalhaes, *"Estereoscopio com tela holografica para ver tomografias",* http://arxiv.org/pdf/1004.2275
11) J.J. Lunazzi, A. Turatto, S.S, Moreira, *"Adaptação de um projetor multimídia a um par estereoscópico por meio de espelhos",* trabalho a publicar.
12) L. P. Yaroslavsky, J. Campos, M. Espínola, I. Ideses, *"Redundancy of stereoscopic images: Experimental evaluation",* Optics Express, Vol. 13, Issue 26, pp. 10895-10907 (2005)    doi:10.1364/OPEX.13.010895
http://www.opticsinfobase.org/viewmedia.cfm?uri=oe-13-26-10895&seq=0
13) J.J. Lunazzi, *"Breve vídeo 3D bicolor",* exemplo no YouTube:
http://www.youtube.com/watch?v=uNJCX5UtDYo
14) J.J. Lunazzi, J.M.J. Ocampo, *"Visualização Tridimensional por Técnicas de Estereoscopia e sua Extensão para a Visualização Holográfica",* trabalho 7/36 na I

Mostra de trabalhos da Unicamp em computação de imagens. Departamento de Engenharia de Computação da Unicamp, 1992.

15) Fábio Barros Cavalcante, exemplo de desenho animado feito por em 1994-5, orientado por J.J. Lunazzi. Quase no final da página:
http://www.ifi.unicamp.br/~lunazzi/prof_lunazzi/Estereoscopia/estere.htm

16) Tutoriais sobre *Como fazer gif animado com o programa livre GIMP*:
http://www.youtube.com/watch?v=4yhc1-l0CIY
http://www.youtube.com/watch?v=mFBmOHRFKnA

17) C. Massao Kawata, J.J. Lunazzi, *"Estereoscopia com espelho e estereogramas",* II Mostra de Trabalhos da UNICAMP em Computação de Imagens, org.: CEPAGRI-CCUEC-DFA/IFGW, Campinas, SP, 09/11/94

18) Usar o buscador Google com a palavra "anaglyph", selecionando a opção imagens. No dia 10 de dezembro de 2010 resultaram 147.000, a cada dia aparecem mais.

19) Artigo prévio de preparação deste: "Fazendo 3D com uma Câmera Só", J.J. Lunazzi http://arxiv.org/pdf/1012.3095